\documentclass{optica-article}

\journal{opticajournal} 

\articletype{Research Article}

\usepackage{physics} 
\usepackage{ulem} 
\usepackage{graphicx} 
\usepackage{subcaption} 

\begin{document}

\title{Ultra-high-rate detection of entangled photon pairs}

\author{Toshimori Honjo,\authormark{1,*} Shigeyuki Miyajima,\authormark{2} Shigehito Miki,\authormark{2} Hirotaka Terai,\authormark{2} Hsin-Pin Lo,\authormark{1} Takuya Ikuta,\authormark{1} Yuya Yonezu,\authormark{1} and Hiroki Takesue,\authormark{1}}

\address{\authormark{1}Basic Research Laboratories, NTT, Inc., 3-1 Morinosato Wakamiya, Atsugi, Kanagawa 243-0198 Japan\\
         \authormark{2}Advanced ICT Research Institute, National Institute of Information and Communications Technology, 588-2 Iwaoka, Nishi-ku, Kobe 651-2492 Japan}

\email{\authormark{*}toshimori.honjo@ntt.com} 


\begin{abstract*} 
The high-rate detection of entangled photons is essential for advancing photonic quantum information processing. Although several experimental demonstrations have been reported, the achievable coincidence rates have so far remained limited. One of the main bottlenecks arises from the dead time of single-photon detectors, which constrains coincidence detection at high photon-pair generation rates.
In this work, we employ 16-pixel superconducting nanowire single-photon detectors (SNSPDs) to mitigate the impact of detector dead time. Consequently, we achieve coincidence rates exceeding 3 million counts per second (Mcps) in two-photon interference and CHSH inequality experiments using 5-GHz clocked sequential time-bin entangled photon pair source. 
To the best of our knowledge, this is the first demonstration of multi-Mcps coincidence detection of entangled photons, paving the way for high-speed entangled-photon-based quantum information processing.

\end{abstract*}


\section{Introduction}\label{sec1}

Entangled photons are a fundamental resource for photonic quantum information processing.
The performance of linear optical quantum computing relies heavily on the ability to detect entangled photons at the output of quantum circuits\cite{Slussarenko_2019}. Similarly, in quantum repeaters, achieving high detection rates at the nodes increases the success rate of the teleportation of the quantum state\cite{RevModPhys.95.045006}. Moreover, in quantum key distribution, higher coincidence detection rates directly lead to increased key generation rates\cite{Gisin_2002}\cite{Yin2020}\cite{Neumann2022}\cite{PhysRevLett.134.230801}.
Over the past several decades, various experiments have successfully demonstrated entangled photon pair distribution.
Recently, high-rate entangled photon pair distribution has gained attention. Several groups have explored wavelength-division multiplexing (WDM) with broadband frequency-correlated entangled photon pair sources. For instance, S. P. Neumann et al. achieved a total coincidence rate of 120 kcps over 7 channels (16 kcps per channel)\cite{Neumann2022experimental}, and D. Aktras et al. reported 38 kcps using 8 channels (4.5 kcps per channel) for time-bin entangled photons\cite{Aktas_2016}. In 2024, A. Mueller et al. reported an experiment with a coincidence rate of 3.55 Mcps across 8 channels (443 kcps per channel)\cite{Mueller:24}. However, their experimentally demonstrated rate was 80 kcps per channel, and the 443 kcps value was estimated under the assumption of using four detectors per a pair of channels. In contrast, K. Wakui et al. reported 1.6 Mcps for polarization-entangled photon-pair detection using three multiplexed superconducting nanowire single-photon detectors (SNSPDs) in 2020 \cite{Wakui:20}. Nevertheless, this value was obtained by summing four independent measurements, and thus the peak coincidence rate is estimated to be approximately 800 kcps.
Despite the development of high-repetition-rate photon pair sources, the straightforward practical coincidence detection rates have not reached the megacount-per-second level.

High-rate coincidence measurements of entangled photon pairs require careful consideration of detector dead time, which has been explicitly addressed in only a limited number of studies. Because of channel losses, including those in the photon source, as well as the finite detection efficiency of the detectors, the detection events of signal and idler photons are inherently probabilistic. If signal photon is detected while the idler's detector is within its dead time, coincidence detection cannot occur; coincidence events are therefore possible only when both detectors are ready. This limitation becomes particularly significant when the detector dead time is longer than the repetition frequency of the photon pair source. The results of a simple numerical simulation are provided in Appendix \ref{secA}.
  Consequently, when using the sources that probabilistically generate entangled photon pairs, achieving high-speed coincidence measurements requires both increasing the repetition rate of the photon source and reducing the detector dead time.

In this work, we employed 16-pixel SNSPDs to overcome the detector dead time limitation\cite{Miki:21}. While a typical SNSPD has a dead time of approximately 50 ns, the multi-pixel architecture allows individual pixels to function independently. Thus, when one pixel detects a photon, the remaining pixels remain active, effectively reducing the overall system dead time. To demonstrate high-rate coincidence detection, we used a 5-GHz-clocked sequential time-bin entangled photon pair source\cite{PhysRevLett.82.2594}, stable planar lightwave circuit Mach-Zehnder interferometers (PLC-MZIs) with 200-ps delay, a pair of 16-pixel SNSPDs and a custom FPGA-based time-to-digital converter (TDC). Using this setup, we achieved a coincidence rate of over 3 Mcps - the highest rate reported to date, to the best of our knowledge.
This technology is expected to enable a significant acceleration of entangled-photon-based quantum information processing.

\newpage
\clearpage

\section{Methods}\label{sec3}

\subsection{Sequential time-bin entangled photon pair}\label{subsec31}

Time-bin entanglement is a form of quantum entanglement in which the quantum states of photons are correlated in the temporal domain. The time-bin scheme is widely adopted in fiber-based quantum communication due to its robustness against fluctuations in the refractive index and birefringence of optical fibers. Specifically, the relative phase between temporal modes in a time-bin qubit remains stable during fiber transmission.

In our experiment, we employed high-dimensional time-bin entanglement - commonly referred to as sequential time-bin entanglement to achieve high coincidence count rates by efficiently utilizing the time domain\cite{PhysRevA.69.050304}\cite{Zhang:08}\cite{Inagaki:13}. This type of entanglement can be generated via spontaneous parametric down-conversion (SPDC) in a nonlinear crystal pumped by temporally separated, coherent sequential pulses. When the average number of photon pairs per pulse is sufficiently low, the whole state wave function can be approximated as:

\begin{equation}
  \ket{\phi} = \frac{1}{\sqrt{N}}\sum_{k=1}^{N} \ket{k}_{s}\ket{k}_{i}
  \label{eq7}
\end{equation}

where $\ket{k}_{z}$ denotes the state in which a single photon is located in the $k$-th time slot of mode $z$ (=s(signal), i(idler)). $N$ is the number of time bins (or pulses) over which the phase coherence of the pump laser is preserved.
After spectral separation, the signal and idler photons are directed to their respective measurement setups. To evaluate the degree of entanglement between signal and idler photons, we perform two-photon interference using a Franson-type interferometer. Each photon is measured using a 1-bit delayed MZI, followed by a single-photon detector. The 1-bit delayed MZI transforms the input state $\ket{k}_{z}$ into an unnormalized superposition $\ket{k}_{z} + e^{i\theta_{z}}\ket{k + 1}_{z}$ where $\theta_{z}$ is the relative phase difference between the two arms of the interferometer. Substituting this transformation into Eq. \ref{eq7}, the entangled state becomes:

\begin{equation}
  \ket{\phi} = \ket{1}_{s}\ket{1}_{i} + \sum_{k=2}^{N-1} (1 + e^{i(\theta_{s}+\theta_{i})})\ket{k}_{s}\ket{k}_{i} + e^{i(\theta_{s}+\theta_{i})}\ket{N}_{s}\ket{N}_{i}
  \label{eq8}
\end{equation}

Here, only the components that contribute to coincidence counts are retained, and normalization is omitted for simplicity. When $N >> 1$, the coincidence count rate is approximately proportional to $1 + V cos(\theta_{s} + \theta_{i})$, where $V$ denotes the two-photon interference visibility. By measuring the coincidence probabilities while varying the phases $\theta_{s}$ and $\theta_{i}$ of the MZIs, we can observe interference fringes that serve as a signature of time-bin entanglement.

\newpage
\clearpage

\subsection{Multi-pixel SNSPD with SFQ circuit}\label{subsec32}

Superconducting nanowire single-photon detectors (SNSPDs) offer high detection efficiency and are widely used in quantum optics experiments. However, conventional SNSPDs typically exhibit a dead time of approximately 50 ns, limiting the maximum achievable count rate to around tens of Mcps. As discussed in the introduction, reducing detector dead time is essential for realizing high-rate coincidence detection.
A promising approach is to use a multi-pixel SNSPD, which consists of multiple SNSPD elements\cite{Rosenberg:13}\cite{doi:10.1021/acsphotonics.4c01680}. This architecture enables photon detection even when some pixels are still recovering from previous events, allowing the system to detect photons arriving at intervals shorter than the dead time of an individual SNSPD.
Note that pixels undergoing dead time are unable to detect photons, resulting in a small reduction in the effective quantum efficiency.
Meanwhile, reading out signals from multiple pixels requires a large number of electrical cables, which imposes a significant heat load on the cryocooler. A solution to this challenge is to incorporate a single-flux quantum (SFQ) circuit, a low-power digital logic circuit optimized for cryogenic operation. By implementing a multiplexing function within the SFQ circuit, we can significantly mitigate the thermal load due to cabling.
After multiplexing the signals from 16 pixels,the SFQ circuit generates electrical pulses with durations of approximately 400 ps at the rising edge of each photon detection event.
In general, crosstalk between pixels can lead to spurious detection events in multi-pixel SNSPD systems. However, in the present system, the signals from multiple pixels are multiplexed in SFQ, and coincidence events within a 400 ps time window are discarded. Therefore, the effect of crosstalk on the measurement results is considered negligible.
By integrating a 16-pixel SNSPD with an SFQ-based readout circuit, we realized an ultra-low-dead-time single-photon detection system using only a single readout line\cite{Miki:21}.

\newpage
\clearpage

\subsection{Experimental setup}\label{subsec33}

\begin{figure}[h]
\centering
\includegraphics[width=1.2\textwidth]{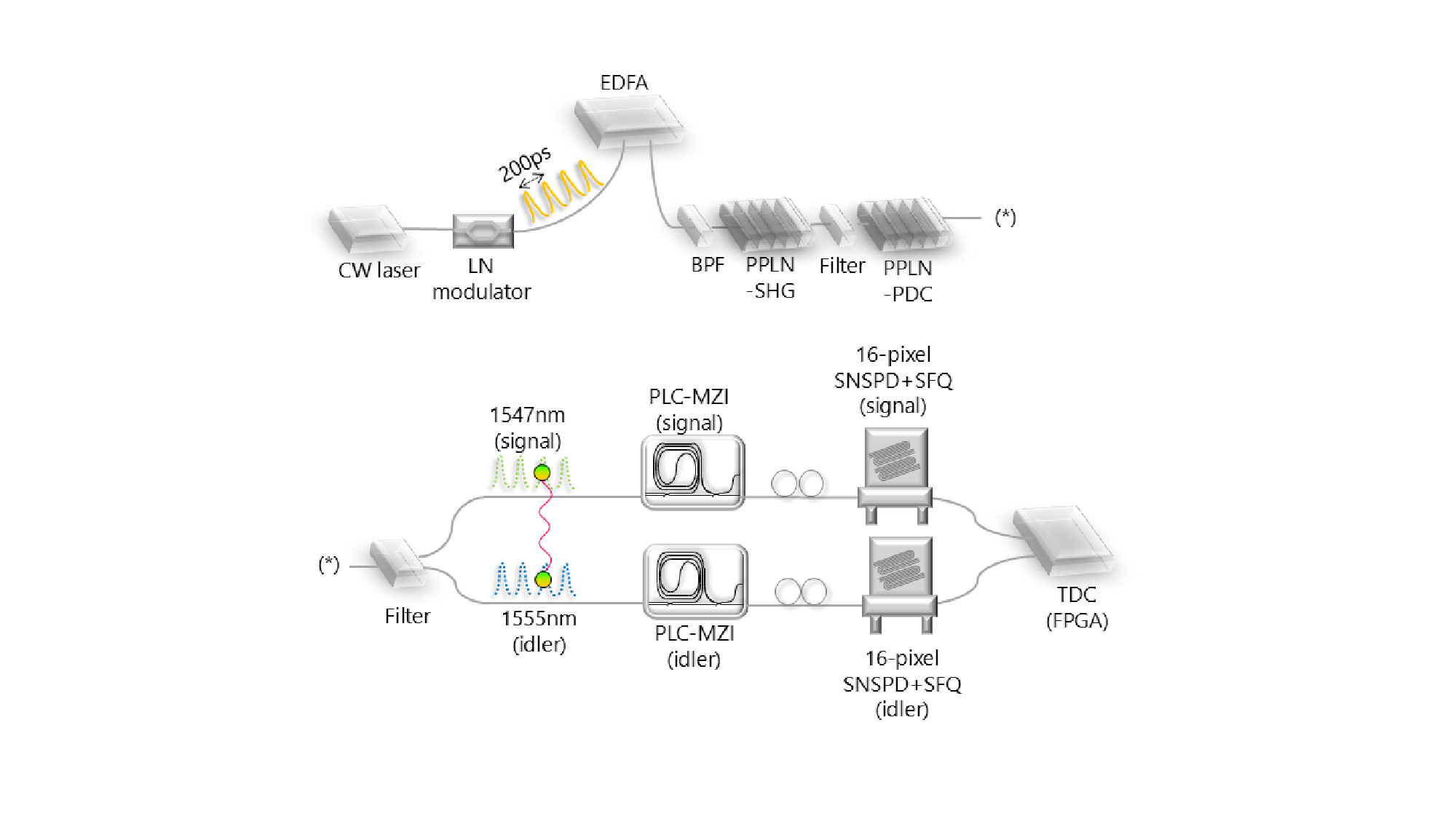}
\caption{Schematic of the experimental setup for generating and detecting sequential time-bin entangled photon pairs: CW laser, continuous-wave laser; LN modulator, $LiNbO_{3}$ modulator; EDFA, erbium-doped fiber amplifier; BPF, band-pass filter; PPLN, periodically poled lithium niobate; PLC-MZI, planar lightwave circuit Mach-Zehnder interferometer; SNSPD, superconducting nanowire single-photon detector; SFQ, single-flux quantum; TDC, time-to-digital converter.}\label{Fig-setup}
\end{figure}

Figure \ref{Fig-setup} shows our experimental setup. A continuous-wave light at a wavelength of 1551 nm, emitted from an external-cavity semiconductor laser, was converted into a 5 GHz pulse train using a $LiNbO_{3}$ intensity modulator. The pulse width was approximately 50 ps. The coherence time of the laser output was about 10 $\mu$s, which is much longer than the temporal interval between pulses. The pulse train was amplified using an erbium-doped fiber amplifier (EDFA) and spectrally filtered by a 1-nm dielectric band-pass filter to suppress amplified spontaneous emission noise. The filtered pulses were directed into a Type-0 periodically poled lithium niobate (PPLN) waveguide (PPLN-SHG), where second harmonic generation produced 775.5 nm sequential pulses. The output from PPLN-SHG was passed through a filter that transmitted the 775.5 nm light while removing residual 1551 nm components. These 775.5 nm pulses were then injected into a second Type-0 PPLN waveguide (PPLN-PDC), where sequential time-bin entangled photon pairs were generated via spontaneous parametric down-conversion (SPDC). The pump, signal, and idler frequencies are denoted by $f_{p}$, $f_{s}$ and $f_{i}$, respectively, and satisfy the relation $f_{p} =f_{s} + f_{i}$. The output from PPLN-PDC was passed through a filter that transmitted the 1551nm photons while removing remaining 775.5nm pump light. A subsequent 1-nm band-pass filter was used to separate the signal (1547nm) and idler (1555nm) photons. These photons were then directed into planar lightwave circuit Mach-Zehnder interferometers (PLC-MZIs) with a path length difference of 4 cm, corresponding to a free spectral range (FSR) of 5 GHz. The phase difference between the interferometer arms was finely and stably controlled via temperature tuning. The output of each PLC-MZI was connected to a 16-pixel SNSPD. The detector signals were multiplexed using a single-flux quantum (SFQ) circuit, enabling ultra-low dead-time operation and high count rates. Finally, the output signals were processed with a custom FPGA-based TDC, in which a digital delay-line (or tapped delay-line) method was implemented on the FPGA evaluation board to record the detection events (see Appendix \ref{secB}).

\newpage
\clearpage


\section{Results}\label{sec2}

\subsection{Coincidence measurement}\label{subsec21}

First, we conducted a coincidence measurement to characterize the basic performance of our entangled photon pair source\cite{PhysRevA.72.041804}\cite{TAKESUE2010276}.
Assuming a low average number of photon pairs per pulse and a low dark-count probability, the average detection probability per pulse for the signal and idler channels is given by:
\begin{equation}
  R_{s} = \mu_{c} \alpha_{s} + d_{s}  \label{eq1}
\end{equation}
\begin{equation}
  R_{i} = \mu_{c} \alpha_{i} + d_{i}  \label{eq2}
\end{equation}
where $\mu_{c}$, $\alpha_{x}$ and $d_{x}$ denote the average number of photon pairs per pulse, the transmittance of channel $x$, and the dark count probability of channel $x$ respectively, with $x = s$ (signal) or $i$ (idler), respectively.
Note that this average number of photon pairs represents the mean number of photons per pulse in each mode of the signal and idler selected by the frequency filter.
In our experimental configuration, the pump pulse duration was much longer than the coherence time of the down-converted photons, so the probability of generating $n$ pairs in a given pulse follows a Poisson distribution. The coincidence and accidental coincidence probabilities are described by:
\begin{equation}
  R_{cc} \simeq \mu_{c} \alpha_{s} \alpha_{i} + (\mu_{c} \alpha_{s} + d_{s})(\mu_{c} \alpha_{i} + d_{i})  \label{eq3}
\end{equation}
\begin{equation}
  R_{acc} \simeq (\mu_{c} \alpha_{s} + d_{s})(\mu_{c} \alpha_{i} + d_{i})  \label{eq4}
\end{equation}
The coincidence-to-accidental ratio (CAR) is then given by:
\begin{equation}
  CAR = \frac{R_{cc}}{R_{acc}} = \frac{\mu_{c} \alpha_{s} \alpha_{i}}{(\mu_{c} \alpha_{s} + d_{s})(\mu_{c} \alpha_{i} + d_{i})} + 1  \label{eq5}
\end{equation}

\begin{figure}[h]
\centering
\includegraphics[width=0.6\textwidth]{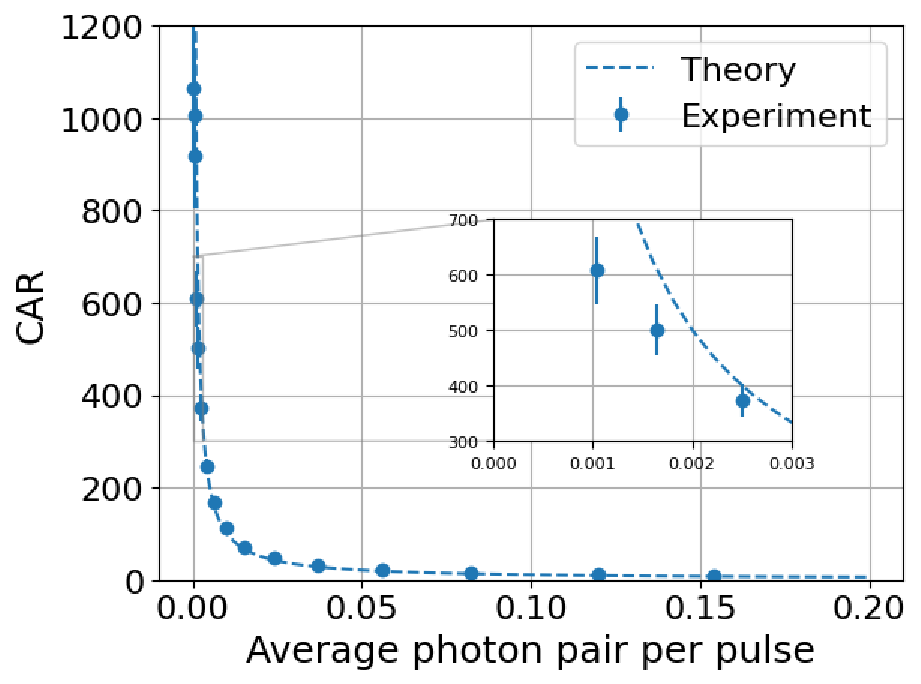}
\caption{Experimentally obtained coincidence-to-accidental ratio (CAR) plotted as a function of the estimated the average number of photon pairs per pulse, $\mu_{c}$. The dashed line indicates the theoretical CAR assuming negligible dark counts. Although each measurement was performed in a single shot, the error bars were estimated from Poissonian counting statistics of the detected counts, and converted into uncertainties of the CARs.
The error bars are small and visible only in the inset.}\label{Fig-car}
\end{figure}

Using the setup shown in Fig. \ref{Fig-setup}, excluding the PLC-MZIs, we measured the CAR and count rates for various pump powers. To estimate $\mu_{c}$ for each pump power, it is necessary to know the transmittance values $\alpha_{s}$ and  $\alpha_{i}$. However, accurately determining the transmittance including excess loss in the periodically poled lithium niobate (PPLN) waveguide and fiber coupling efficiency is challenging. Therefore, we selected values for $\alpha_{s}$ and  $\alpha_{i}$ that best fit the measured CAR and count rates for the estimated $\mu_{c}$. Taking into account the dead time (2ns) of the TDC, the single count rate is described by \cite{Takesue_2005}:
\begin{equation}
  CR_{signal} = f \mu_{c}\alpha_{s} e^{-f \mu_{c}\alpha_{s} t_{d}} \label{eq5-1}
\end{equation}
where $f$ and $t_{d}$ denotes the clock frequency and the dead time of the TDC. Note that pixels in dead time cannot detect photons, resulting in a slight reduction in the effective quantum efficiency. This reduction, however, is not taken into account in the theoretical model described above.
Using Eq. \ref{eq5-1}, $\mu_{c}$ was estimated from the single count rate.
Figure \ref{Fig-car} shows the experimentally obtained CAR as a function of $\mu_{c}$. The detection efficiency of the SNSPDs was measured in advance to be 67\% (see Appendix \ref{secC}).
The estimated transmittances $\alpha_{s}$ and $\alpha_{i}$, excluding the detection efficiency of the SNSPDs, were both approximately $-$3.5 dB. The dashed line in Fig. \ref{Fig-car} indicates the theoretical CAR assuming negligible small dark counts. These results confirm the fundamental performance of our entangled photon pair source.

\newpage
\clearpage

\subsection{Two photon interference experiment}\label{subsec22}

Next, we performed a two-photon interference experiment to evaluate the quality of the distributed entangled photon pairs at various coincidence rates\cite{Inagaki:13}\cite{PhysRevA.72.041804}\cite{Honjo:07}. In this experiment, we measured coincidence counts while keeping the temperature of the PLC-MZI in the signal channel fixed and varying the temperature of the PLC-MZI in the idler channel. Changing the temperature corresponds to varying the phase difference in the interferometer.

Initially, we fixed the signal-side interferometer phase $\theta_{s}$  by setting the temperature of the signal-channel PLC-MZI to 40.0${}^\circ$C. We then swept the temperature of the idler-channel PLC-MZI from 43.0${}^\circ$C to 46.0${}^\circ$C in 0.1${}^\circ$C steps while recording the coincidence counts. For each temperature setting, a total of $2.62 \times 10^{5}$ events were collected for both the signal and idler channels. This experiment was repeated for several values of the average number of photon pairs per pulse by adjusting the pump power. 

\begin{figure}[htbp]
  \centering

  \begin{subfigure}{0.45\textwidth}
    \centering
    \includegraphics[width=\linewidth]{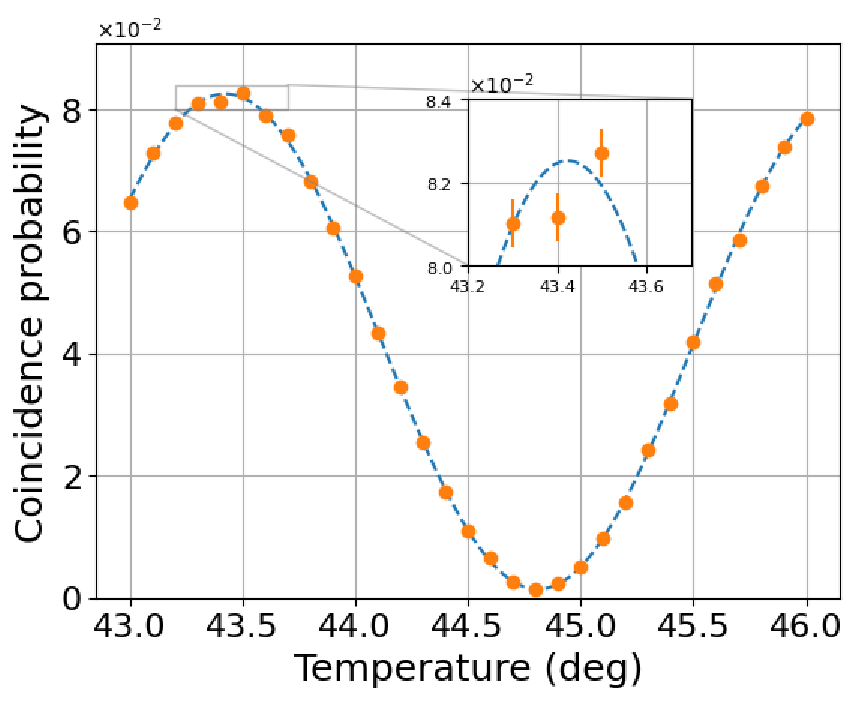}
    \caption{}
  \end{subfigure}
  \hfill
  \begin{subfigure}{0.45\textwidth}
    \centering
    \includegraphics[width=\linewidth]{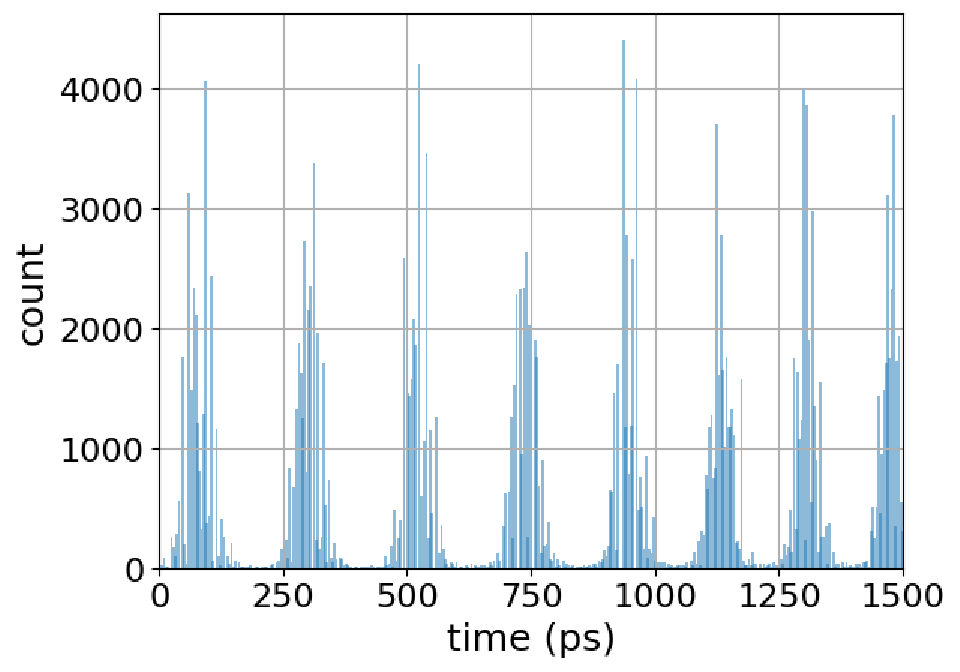}
    \caption{}
  \end{subfigure}

  \vspace{0.5cm}

  \begin{subfigure}{0.45\textwidth}
    \centering
    \includegraphics[width=\linewidth]{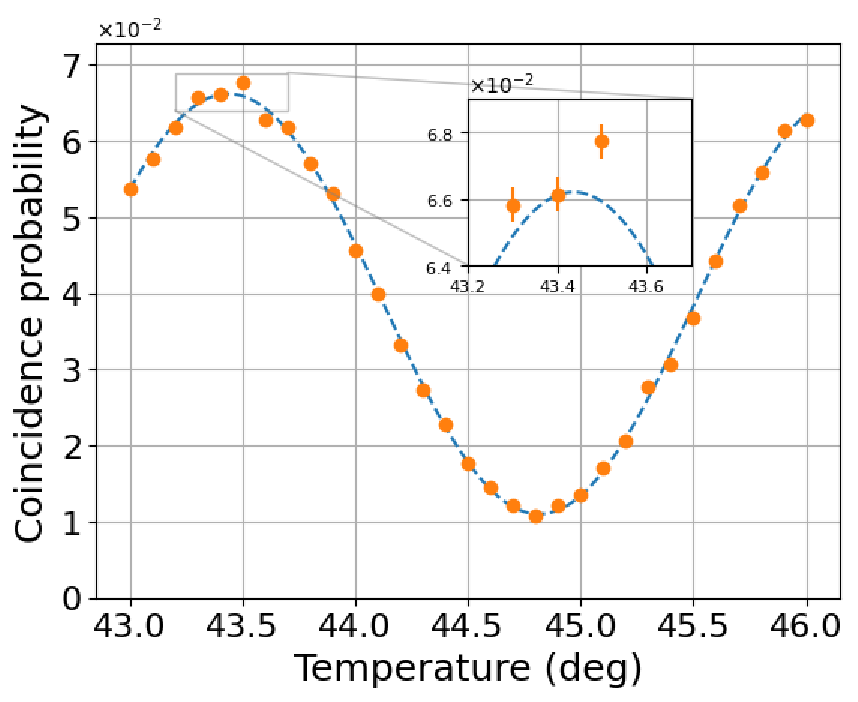}
    \caption{}
  \end{subfigure}
  \hfill
  \begin{subfigure}{0.45\textwidth}
    \centering
    \includegraphics[width=\linewidth]{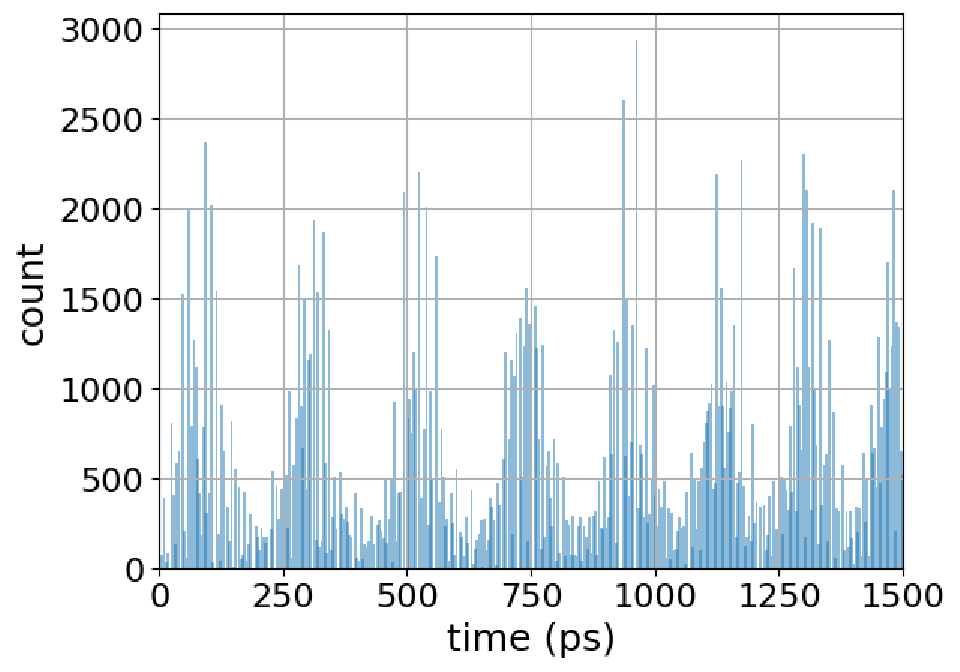}
    \caption{}
  \end{subfigure}

  \caption{(a) Two-photon interference fringes measured at a low average photon-pair rate ($\mu_{c} = 0.001$). The measured visibility was 96.6\%, and the peak coincidence and signal count rates were 47 kcps and 532 kcps, respectively. (b) Corresponding histogram of single-photon detection events in the signal channel at a low average photon-pair rate. (c) Two-photon interference fringes measured at a high average photon-pair rate ($\mu_{c} = 0.094$). The visibility was 71.4\%, with a peak coincidence rate of 3.3 Mcps and a signal count rate of 47 Mcps. (d) Corresponding histogram of single-photon detection events in the signal channel at a high average photon-pair rate.}\label{Fig-fringe} 
\end{figure}

Figure \ref{Fig-fringe} shows two-photon interference fringes and histograms of single-photon detection events at low and high count rates.
Although the coincidence probability at each temperature was obtained from a single-shot measurement, the error bars were evaluated based on the Poissonian counting statistics of the detected counts. However, because the error bars are very small, they are visible only in the inset.
At a low count rate, where the average number of photon pairs per pulse was 0.001, the visibility reached 96.6\%, with a peak coincidence rate of 47 kcps and a signal count rate of $\sim$532 kcps.
In contrast, at a high count rate, where the average number of photon pairs per pulse was 0.094, the visibility decreased to 71.4\%, while the peak coincidence rate and signal count rate increased to 3.3 Mcps and $\sim$47 Mcps, respectively.
Notably, the visibility remains above the threshold of 70.7\% required for violating Bell's inequality, even at high count rates exceeding 3 Mcps.

\newpage
\clearpage

\begin{figure}[htbp]
  \centering

  \begin{subfigure}{0.6\textwidth}
    \centering
    \includegraphics[width=\linewidth]{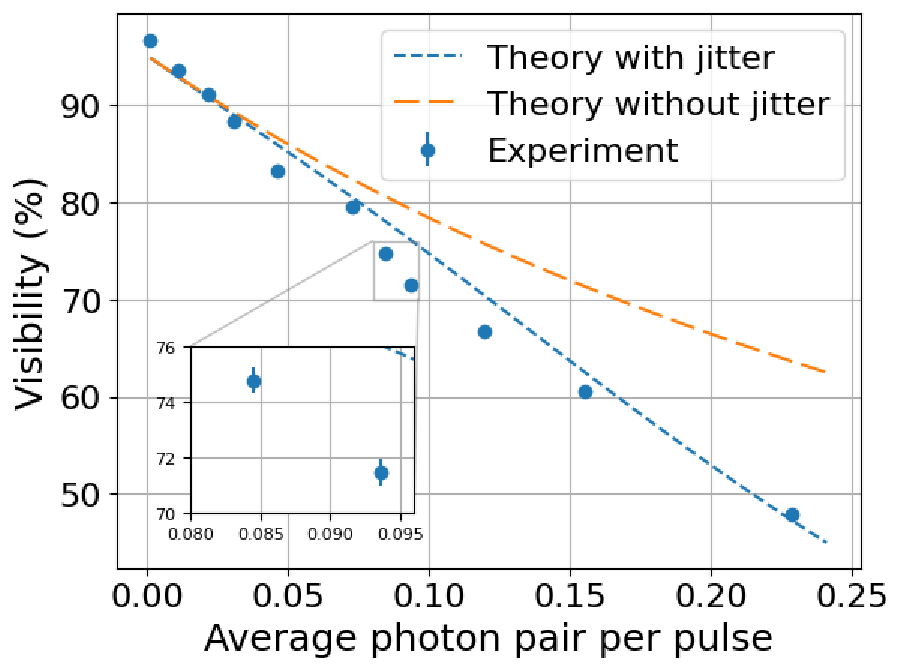}
    \caption{}
  \end{subfigure}

  \vspace{0.5cm}

  \begin{subfigure}{0.45\textwidth}
    \centering
    \includegraphics[width=\linewidth]{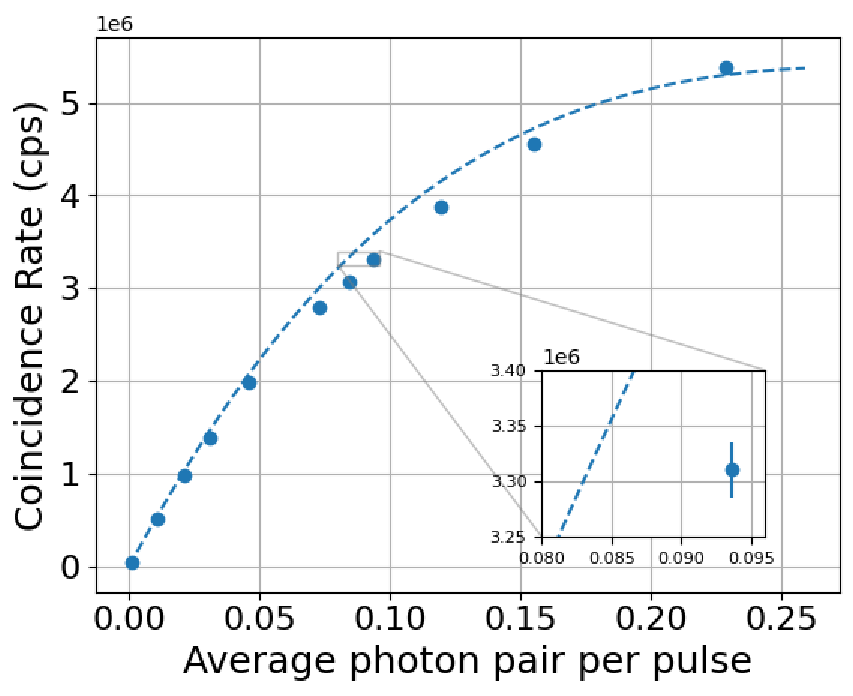}
    \caption{}
  \end{subfigure}
  \hfill
  \begin{subfigure}{0.45\textwidth}
    \centering
    \includegraphics[width=\linewidth]{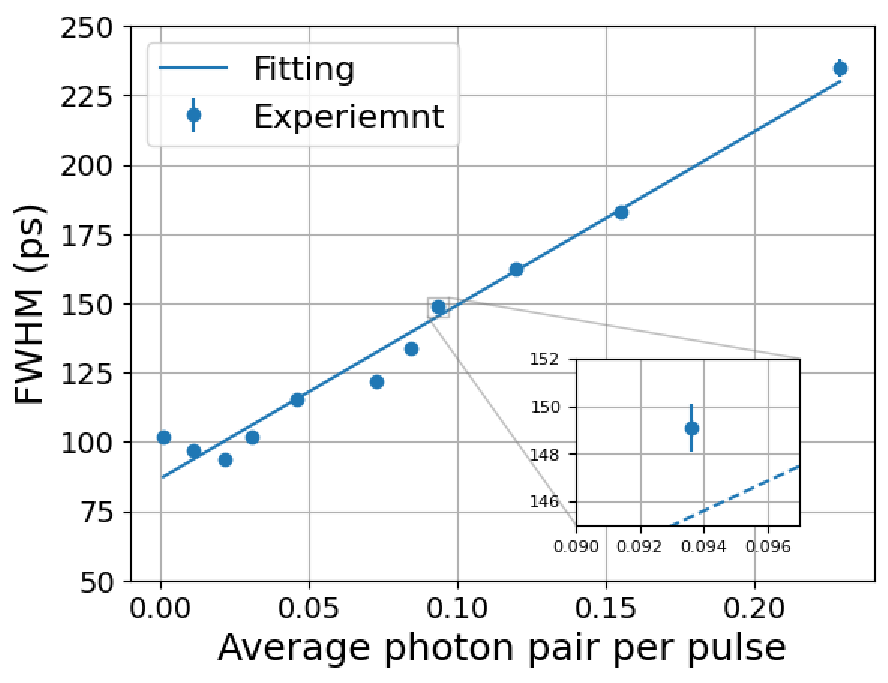}
    \caption{}
  \end{subfigure}

  \caption{(a) Visibility of two-photon interference experiment as functions of the average number of photon pairs per pulse. (b) Coincidence rate as functions of the average number of photon pairs per pulse. The dashed line shows the theoretically estimated coincidence rate. (c) The jitter of FWHM as a cuntion of the average number of photon pairs per pulse. The solid line represents linear fits to the data. For panels (a) and (b), the error bars were estimated from the Poissonian counting statistics of the detected counts, whereas for panel (c), they represent the fitting errors estimated from a fit that accounts for Poissonian counting statistics; in all panels, the error bars are small and visible only in the inset.}\label{Fig-TPI}

\end{figure}

Figure \ref{Fig-TPI} (a) shows the visibility and Fig. \ref{Fig-TPI} (b) shows the coincidence rate as functions of the average number of photon pairs per pulse. The degradation in visibility is primarily attributed to accidental coincidences arising from multi-pair emission and the timing jitter of the single-photon detectors.  As the count rate increases, the temporal overlap of detection events becomes less precise due to the increased jitter.

Here we theoretically analyze the degradation in visibility caused by multi-pair emission and the timing jitter of the single photon detectors. To assess the impact of timing jitter on this degradation, we quantified its contribution separately.
First, we fitted the histogram of single-photon detection events in the two-photon interference experiment with a Gaussian function to estimate the full width at half maximum (FWHM) at each pump power. Figure \ref{Fig-TPI} (c) shows the FWHM of the timing jitter as a function of the average number of photon pairs per pulse.
From the value of the FWHM, we can estimate the detection probability $P_{in}$, originating from the target pulse, and $P_{out}$, originating from its adjacent pulse. We also note that, in the following discussion, we assume $P_{in} + P_{out} \simeq 1$.
In the following, we assume that dark counts are neglected for simplicity, and we also omit the channel transmittance, as it does not affect the visibility. The coincidence probability $R_{cc}$, where signal and idler photons from the target pulse are simultaneously detected within a given time slot, is expressed as
\begin{equation}
  R_{cc} \simeq \mu_{c} P_{in}P_{in} + R_{acc}.
\end{equation}
For the accidental coincidence probability $R_{acc}$, we can consider four cases arising from different photon pairs:
\begin{enumerate}
\item Signal and idler photons both from the target pulse: $\mu_{c} P_{in} \cdot \mu_{c} P_{in}$
\item Signal photon from the target pulse, idler photon from the adjacent pulse: $\mu_{c} P_{in} \cdot \mu_{c} P_{out}$
\item Signal photon from the adjacent pulse, idler photon from the target pulse: $\mu_{c} P_{out} \cdot \mu_{c} P_{in}$
\item Signal and idler photons both from the adjacent pulse: $\mu_{c} P_{out} \cdot \mu_{c} P_{out}$
\end{enumerate}
The accidental coincidence probability $R_{acc}$ is expressed as
\begin{equation}
  R_{acc} \simeq \mu_{c}^2 P_{in}(P_{in} + P_{out}) + \mu_{c}^2 P_{out} (P_{in} + P_{out}) = \mu_{c}^2.
\end{equation}
Based on these considerations, the visibility can be estimated as follows.
\begin{equation}
  V_{multi-photon\ and\ jitter}= \frac{R_{cc}-R_{acc}}{R_{cc}+R_{acc}} = \frac{P_{in}^2}{P_{in}^2 + 2\mu_{c}} \label{eqB1}
\end{equation}
When the effect of jitter is neglible small, the visibilty is estimated as
\begin{equation}
  V_{multi-photon} = \frac{1}{1+2\mu_{c}}. \label{eqB2}
\end{equation}

Additionally, we included a fixed 5\% error arising from imperfect interference in the MZI.
Figure \ref{Fig-TPI} (a) presents the estimated visibility together with the experimental results.
The blue and orange dashed line shows the estimated visiblity based on Eq. \ref{eqB1}, and \ref{eqB2}, respectively.
The estimated values shows agreement with the measured data.
In addition, by comparing the orange and blue lines, one can clearly observe the effect of timing jitter.
As shown, at higher values of $\mu_{c}$, the influence of jitter becomes more pronounced.


\newpage
\clearpage

\subsection{CHSH inequality experiment}\label{subsec23}

Finally, we performed a CHSH (Clauser-Horne-Shimony-Holt) inequality test\cite{PhysRevLett.23.880}. Demonstrating the violation of Bell's inequality in the CHSH form for two-qubit systems is a widely accepted method to confirm the presence of quantum entanglement. The CHSH parameter $S$ is calculated as follows:
\begin{equation}
  S = E(d_{s}, d_{i}) + E(d_{s}, d_{i}') + E(d_{s}', d_{i}) - E(d_{s}', d_{i}')
  \label{eq5}
\end{equation}
Here, $d_{z}$, $d_{z}'$ (z = s, i) represent specific phase settings $\theta_{z}$ of the interferometers. The correlation function $E(\theta_{s}, \theta_{i})$ is defined as:
\begin{equation}
  E(\theta_{s}, \theta_{i})=\frac{R(\theta_{s}, \theta_{i}) - R(\theta_{s}, \theta_{i}+\pi) - R(\theta_{s}+\pi, \theta_{i}) + R(\theta_{s}+\pi, \theta_{i}+\pi)}{R(\theta_{s}, \theta_{i}) + R(\theta_{s}, \theta_{i}+\pi) + R(\theta_{s}+\pi, \theta_{i}) + R(\theta_{s}+\pi, \theta_{i}+\pi)}
  \label{eq6}
\end{equation}
In this expression, $R(\theta_{s}, \theta_{i})$  denotes the coincidence count rate when the interferometer phases for the signal and idler photons are set to $\theta_{s}$ and $\theta_{i}$, respectively. A measured value of $|S| > 2$ indicates a violation of Bell's inequality. Quantum mechanics predicts a maximum value of $|S| = 2\sqrt{2} \approx 2.828$.

\begin{figure}[h]
\centering
\includegraphics[width=0.6\textwidth]{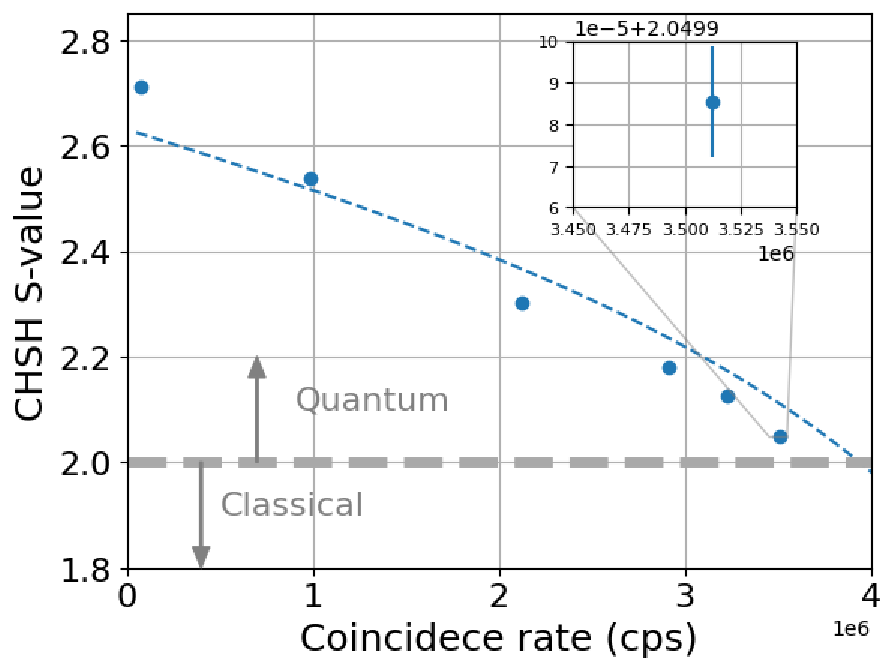}
\caption{CHSH $S$-values plotted as a function of the coincidence count rate. The values above $S=2.0$ indicate a violation of Bell's inequality. Even at rates exceeding 3 Mcps, the measured $S$-values remain well above the classical bound, demonstrating preserved quantum entanglement in time-bin photon pairs.}\label{Fig-chsh}
\end{figure}

We first calibrated the interferometers by setting the base phase values to $\theta_{s0}$ to 40.00${}^\circ$C and $\theta_{i0}$ to 43.40${}^\circ$C, corresponding to the peak of the two-photon interference fringe. From the previously observed two-photon interference fringes, we determined that phase shifts of $\pi$, $\pi/2$ and $\pi/4$ correspond to temperature changes of +1.40${}^\circ$C, +0.7${}^\circ$C, and +0.35${}^\circ$C, respectively.
Using these values, the phase settings for Eq. \ref{eq5} were selected as: $d_{s} = \theta_{s0}$, $d_{s}' = \theta_{s0} + \pi/2$, $d_{i} = \theta_{i0} + \pi/4$, and $d_{i}' = \theta_{i0}-\pi/4$.
We measured all 16 required coincidence rates $R(\theta_{s}, \theta_{i})$ for several values of the average number of photon pairs per pulse. In each setting, we accumulated $2.62 \times 10^{5}$ detection events five times for both the signal and idler channels. Figure \ref{Fig-chsh} shows the measured CHSH $S$-values as a function of the coincidence rate.
The uncertainty of the CHSH $S$-value was evaluated by assuming Poissonian counting statistics for the coincidence counts and propagating these uncertainties through the four correlation coefficients used to calculate $S$.
The dashed line shows the theoretically estimated CHSH $S$-values.
At low pump power, we obtained $S=2.71 \pm 2.12\times10^{-6}$, leading to a violation by 334,905 standard deviations, with a coincidence rate of approximately 76 kcps. At high pump power, we obtained $S=2.05 \pm 1.33\times10^{-5}$, leading to a violation by 3759 standard deviations, with a coincidence rate of approximately 3.5 Mcps. These results demonstrate the successful high-speed distribution and detection of time-bin entangled photon pairs. To the best of our knowledge, this represents the first CHSH inequality test performed with time-bin entangled photon-pair coincidence detection at multi-Mcps rates.

\newpage
\clearpage


\section{Discussion}\label{sec4}

We demonstrated high-rate detection of time-bin entangled photon pairs. Although our system achieves a significant improvement over previous works, the coincidence rate remains slower compared to conventional one-way, weak cohererent laser based quantum communication systems. To further enhance the performance, several improvements can be considered.

A straightforward approach is to reduce the timing jitter of the single-photon detectors. Since the SNSPDs used in this work exhibit a timing jitter of approximately 30 ps, it is in principle possible to operate with a pulse repetition rate of 10-20 GHz.
The SFQ circuit itself can support frequencies exceeding 10-20 GHz\cite{1363670318436564608}.
However, the SFQ circuit design for generating the output signal and the current low-noise amplifier used to boost the output signal from the SFQ circuit cannot sustain such high count rates. A possible solution is to upgrade these design and components to handle faster signals.
Another promising improvement is the incorporation of a constant fraction discriminator (CFD) into the time interval analyzer (TIA)\cite{Chen:13}. At present, we use simple threshold-based discrimination via a comparator. In the high-count-rate regime, the amplitude of the electrical signals from the SNSPD system fluctuates due to bandwidth limitations in the readout circuit. This leads to timing uncertainty known as the ``time walk'' effect. CFDs can mitigate this issue by providing amplitude-independent timing discrimination, improving the overall timing resolution and accuracy of detection. Other approaches have also been proposed, such as calibration methods that exploit the correlation between detection-event delays and the elapsed time between pulses\cite{10.1063/5.0129147}.

Based on the analysis described in Section \ref{subsec22}, if the jitter is completely suppressed, the average number of photon pairs per pulse can be increased to 0.17, achieving a visibility greater than 70.7\%. However, this corresponds to only a factor of 1.8, resulting in an estimated coincidence count rate of 5.76 Mcps. The accidental coincidences arising from multi-photon emissions limit this rate. To further enhance the coincidence rate, it is necessary to increase the clock frequency of the photon-pair source. Furthermore, if the dead time were completely negligible, the saturation factor in Eq. \ref{eq5-1} would be eliminated. In that case, the coincidence rate in Fig. \ref{Fig-TPI}(b) would increase linearly and is expected to reach 4.3 Mcps for $\mu_{c} =0.094$.

On the other hand, enhancing the quality of the photon-pair source is also an important direction. Entangled photon-pair sources based on SPDC in PPLN have been widely studied owing to their high efficiency and compatibility with telecom wavelengths\cite{Zhang:08}\cite{https://doi.org/10.1002/lpor.201400404}\cite{Jin2014}.
Using a shorter PPLN waveguide for SPDC could reduce excess optical loss, provided that sufficient conversion efficiency is maintained. In our setup, the efficiency is already high enough to allow this optimization. Additionally, improving the fiber coupling efficiency would further increase the detected count rates. Looking ahead, the use of deterministic photon-pair sources could enable even more efficient entangled photon distribution, opening new possibilities for scalable quantum information processing.

  Finally, we note the significance of our custom-built FPGA-based time-to-digital converter (TDC). FPGA-based TDCs have attracted attention due to their low implementation cost and suitability for post-processing high-bandwidth data streams. In this work, we used the FPGA only for high-bandwidth digitization and memory storage. In the future, by embedding post-processing functionality into the FPGA, real-time feedback based on photon-counting events becomes feasible \cite{8966320}. Such capabilities may be particularly useful in advanced quantum protocols, including quantum teleportation experiments and photonic quantum information processing.

\newpage
\clearpage

\section{Summnary}\label{sec5}
In summary, we have demonstrated high-rate detection of time-bin entangled photon pairs. By employing 16-pixel superconducting nanowire single-photon detectors (SNSPDs), we achieved a coincidence rate exceeding 3 Mcps in two-photon interference and CHSH inequality experiments using 5-GHz clocked sequential time-bin entangled photon pairs. The two-photon interference visibility was measured to be 71.4\%, and the CHSH inequality yielded a value of $S=2.05 \pm 1.33\times10^{-5}$, corresponding to a violation by 3759 standard deviations. To the best of our knowledge, this represents the first demonstration of time-bin entangled photon-pair detection at multi-Mcps rates. We believe that this achievement constitutes a significant step toward advancing photonic quantum information processing.

\newpage
\clearpage

\appendix

\section{Impact of detector dead time on coincidence measurements}\label{secA}

\begin{figure}[h]
\centering
\includegraphics[width=0.6\textwidth]{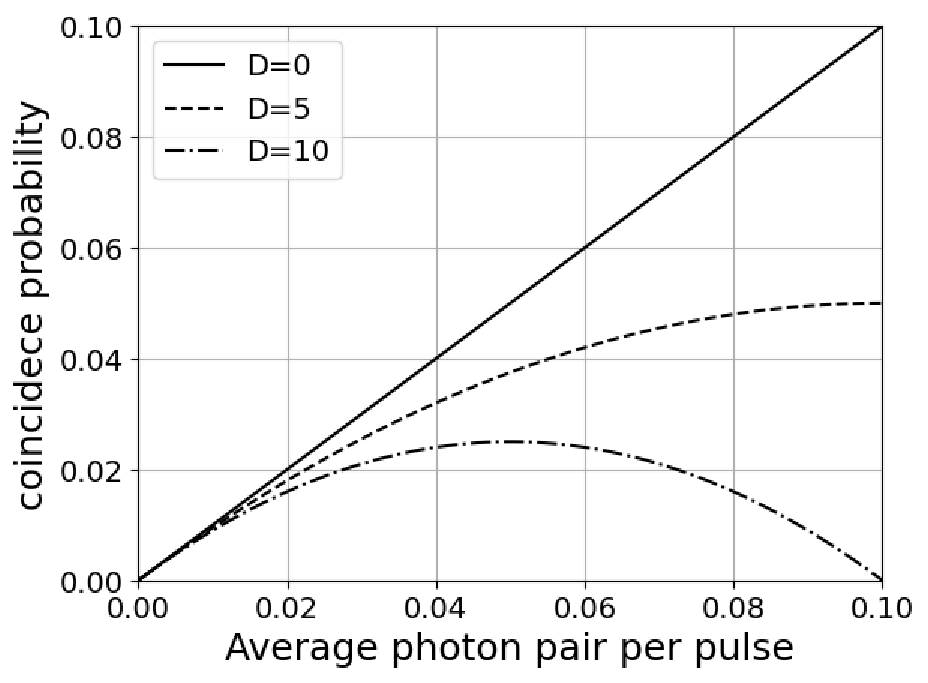}
\caption{Coincidence probability per pulse as a function of average photon pair per pulse. $D$ is the dead time divided by the interval of the pulses.}\label{Fig-conin-dead}
\end{figure}

In the introduction, we described dead time of the single photon detector as a major bottleneck in high-rate coincidence measurements of entangled photon pairs. Here, we estimate the impact of the detector dead time using a simple numerical estimation.
First, we define $D$ as the length of the dead time. Specifically, $D$ is the dead time normalized by the pump pulse interval. In other words, when the pulse interval is fixed, $D$ is given by the dead time divided by the pulse interval.
  For simplicity, we assume a photon-pair source governed by a Poisson distribution, a lossless transmission channel, and detectors with 100\% quantum efficiency. Under these assumptions, all generated photon pairs are guaranteed to reach the detectors. Therefore, the probability of coincident detection in a given time slot, taking into account the dead time, is given by the product of the probability that no photon has been detected in the preceding slots (i.e., that the system is not within the dead time),$1-\mu_{c} D$, and the probability that a photon pair arrives in that slot,$\mu_{c}$. 
Figure \ref{Fig-conin-dead} shows the simulated coincidence probability per pulse as a function of the mean number of photon pairs per pulse, calculated as $\mu_{c}(1-\mu_{c} D)$, where $\mu_{c}$ is the average number of photon pairs per pulse.
The coincidence rate scales linearly with $\mu_{c}$ in the absence of dead time. However, when the dead time equals the duration of five pulses, saturation occurs at $\mu_{c} = 0.1$, yielding only half the coincidence rate compared to the ideal case. For a dead time spanning ten pulses, $\mu_{c}$ must remain below 0.05 to avoid saturation. In practice, losses and detector inefficiencies must also be considered, but the qualitative trend remains unchanged.

\newpage
\clearpage

\section{FPGA-based TDC}\label{secB}

\begin{figure}[h]
\centering
\includegraphics[width=1.0\textwidth]{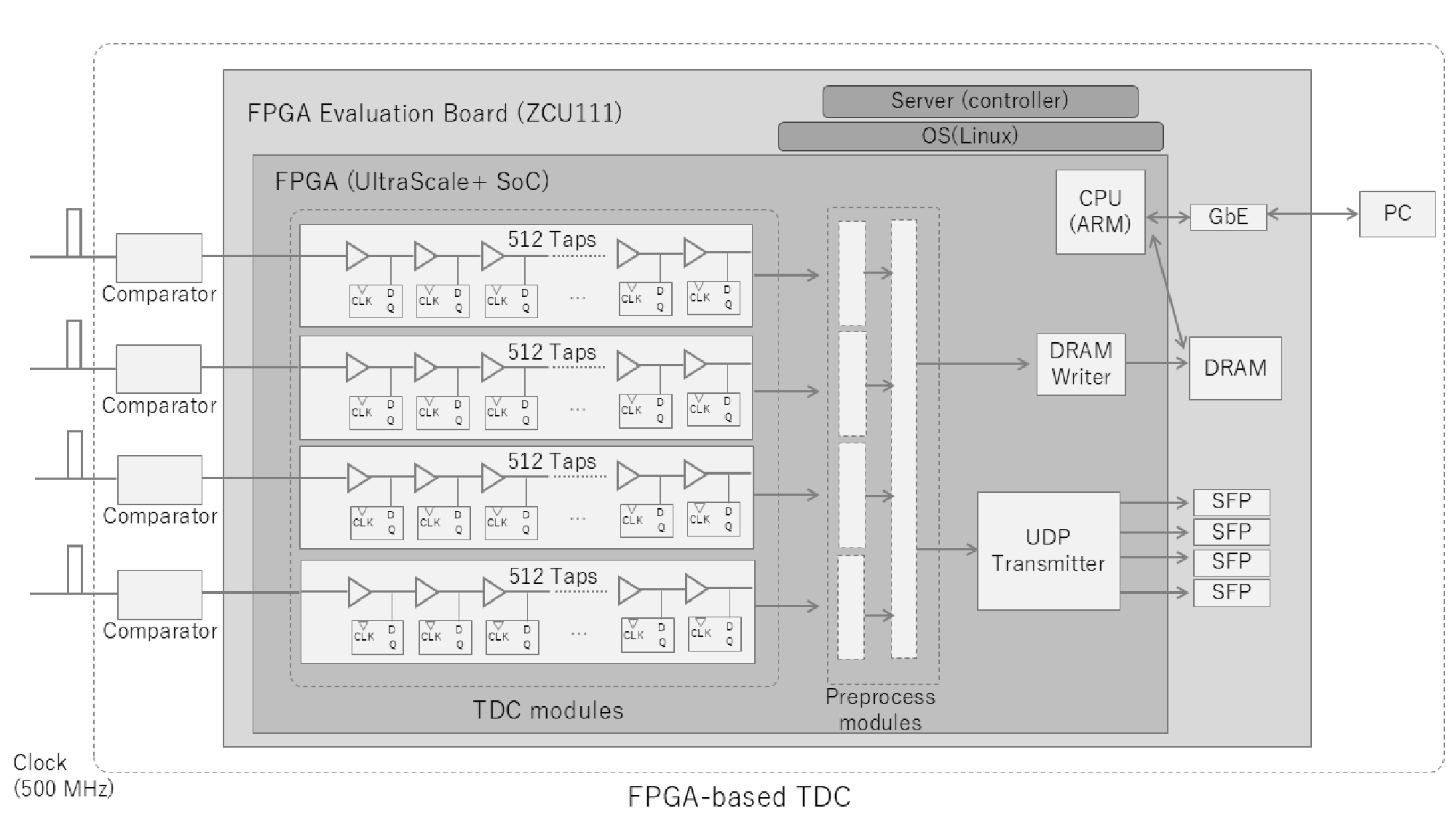}
\caption{Schematic diagram of the FPGA-based TDC.}\label{Fig-FPGA}
\end{figure}

Time-to-digital conversion (or time interval analysis) is one of the main components for achieving a high coincidence count rate. Although several commercial products are available, FPGA-based TDCs have been attracting attention due to their low implementation cost and suitability for post-processing high-bandwidth data streams. With the miniaturization of semiconductors, the time resolution of TDCs has also improved\cite{ZHOU2023168366}\cite{7100940}. Recently, FPGA-based TDCs have been applied in several quantum communication experiments.
The digital delay-line (or tapped delay-line) method is simple and widely used\cite{Kalisz_2004}. The line consists of a chain of cells with well-defined delay times; the input signal propagates through this chain and is successively delayed by each cell. By measuring the number of cells the signal has propagated through at the rising edge of the clock, we can determine the relative input time to the clock. In a simple implementation, an additional clock cycle is necessary to refresh the line after detection. To avoid this additional dead time in the TDC, we applied a dual-mode method, which propagates 1s and 0s in alternating measurement cycles\cite{ZHOU2023168366}\cite{9654192}.
Figure \ref{Fig-FPGA} shows the schematic diagram of our FPGA-based TDC. We implemented a 4-channel TDC on a Xilinx Zynq UltraScale+ RFSoC ZCU111 evaluation board. The 512-tap digital delay line was implemented using CARRY8 logic, and the clock for the TDC module operated at 500 MHz, resulting in a dead time of less than 2 ns. The digitized time data is either stored in DRAM (4GB) or streamed to the SFP interface (10 GbE) for each channel. Our TDC achieves a root-mean-square (RMS) timing resolution of 7 ps and a maximum count rate of 500 Mcps per channel.

\newpage
\clearpage

\section{Performance of 16-pixel SNSPDs}\label{secC}

\begin{figure}[h]
\centering
\includegraphics[width=0.6\textwidth]{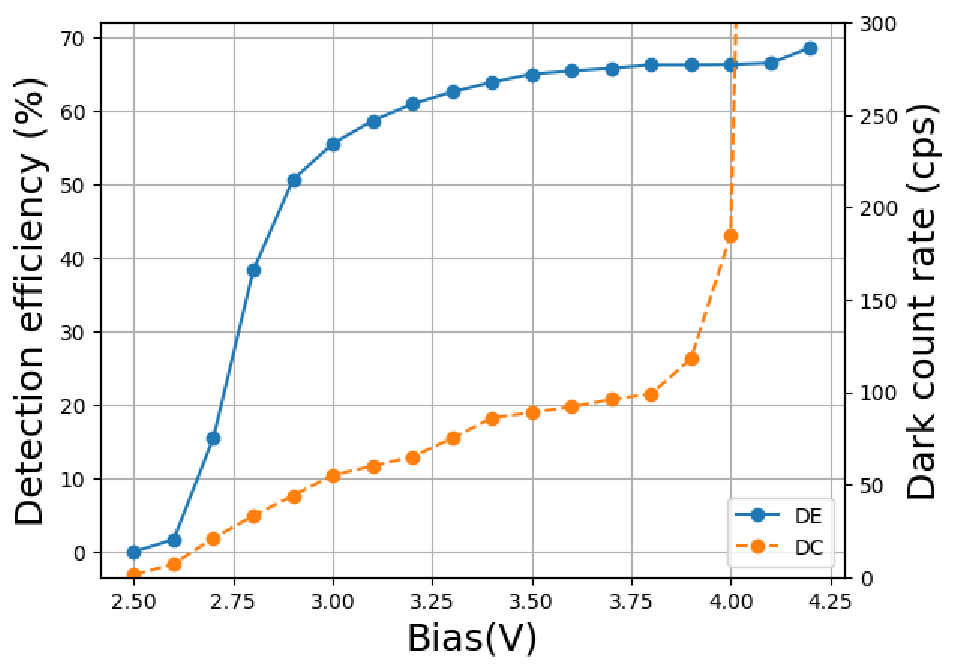}
\caption{Performance of 16-pixel SNSPDs. Detection efficiency(DE) and dark count rate(DC) as functions of the bias voltage.}\label{Fig-sspd-qe-dc}
\end{figure}

The 16-pixel superconducting nanowire single-photon detectors (SNSPDs) are key devices in this experiment. Before performing the photon-pair distribution measurements, we measured the detection efficiency and dark count rate as functions of the bias voltage. To evaluate the detection efficiency, we injected weak coherent pulses at a wavelength of 1551.1 nm, generated using the same setup shown in Fig. \ref{Fig-setup}. The repetition rate was set to be 500 MHz, and the input power was calibrated to $-$100 dBm, corresponding to 780 thousand photons per second. The detection efficiency was estimated from the average detected count rate. Figure \ref{Fig-sspd-qe-dc} shows the detection efficiency and dark count rate as functions of the bias voltage. The bias dependence of the quantum efficiency exhibits a gradual response, which is a result of our device design.

\newpage
\clearpage


\begin{backmatter}

\bmsection{Funding}\label{sec6}
MIC R\&D of ICT Priority Technology (JPMI00316) of Ministry of Internal Affairs Communications of Japan.

\bmsection{Disclosures}\label{sec7}
The authors declare no conflicts of interest.

\bmsection{Data availability}\label{sec8}
Data underlying the results presented in this paper are not publicly available at this time but may
be obtained from the authors upon reasonable request.

\end{backmatter}


\bibliography{3Mcps}

\end{document}